\documentclass[pre,twocolumn, amsmath, amssymb, superscriptaddress]{revtex4}

\usepackage{braket}
\usepackage{subfigure}
\usepackage{adjustbox}
\usepackage{xcolor}
\usepackage{todonotes}
\usepackage{hyperref}

\begin{document}

\title{Simulated Epidemics in 3D Protein Structures to Detect Functional Properties}

\author{Mattia Miotto\footnote{The authors contributed equally to the present work.}}
\affiliation{Department of Physics, Sapienza University of Rome, Rome, Italy}
\affiliation{Center for Life Nanoscience, Istituto Italiano di Tecnologia,  Rome, Italy}

\author{Lorenzo Di Rienzo\footnotemark[1]}
\affiliation{Department of Physics, Sapienza University of Rome, Rome, Italy}

\author{Pietro Corsi}
\affiliation{University ``Roma Tre'', Department of Science, Rome, Italy}

\author{Giancarlo Ruocco}
\affiliation{Department of Physics, Sapienza University of Rome, Rome, Italy}
\affiliation{Center for Life Nanoscience, Istituto Italiano di Tecnologia,  Rome, Italy}

\author{Domenico Raimondo}
\affiliation{Department of Molecular Medicine, Sapienza University of Rome, Rome, Italy}

\author{Edoardo Milanetti}
\affiliation{Department of Physics, Sapienza University of Rome, Rome, Italy}
\affiliation{Center for Life Nanoscience, Istituto Italiano di Tecnologia,  Rome, Italy}


\newcommand{\beq}{\begin{equation}}
\newcommand{\eeq}{\end{equation}}
\newcommand{\red}[1]{\textcolor{red}{{\bf #1}}}

\begin{abstract}
The outcome of an epidemic is closely related to the network of interactions between individuals.
Likewise, protein functions depend on the 3D arrangement of their residues and the underlying energetic interaction network.
Borrowing ideas from the theoretical framework that has been developed to address the spreading of real diseases, we study the diffusion of a fictitious epidemic inside the protein non-bonded interaction network, aiming to study the features of the network connectivity and properties.
Our approach allowed us to probe the overall stability and the capability to propagate information in the complex 3D-structures and proved to be very efficient in addressing different problems, from the assessment of thermal stability to the identification of functional sites.

\end{abstract}

\flushbottom
\maketitle

\thispagestyle{empty}


\section*{Introduction}
Proteins are large bio-molecules responsible for the majority of live-sustaining tasks in cells~\cite{Mannige2014,Chothia1997}. Their great versatility is due to the complex three-dimensional structure they can acquire, which arises as a result of physical and chemical interactions among all its constituent amino acids.
In particular, the global structure is uniquely defined once the sequence of amino acids composing the molecule is specified~\cite{Dill2008}, with different sequences that can give, up to local rearrangements, the same overall 3D architecture~\cite{Lesk1980,Ofran2006}. 

The peculiar structural conformation each protein assumes is the result of a long evolutionary optimization~\cite{Debs2013}. Proteins are adapted to carry on specific tasks, usually binding to other molecules while being embedded in a complex dynamical environment in the presence of both thermal and molecular noises. In this scenario what evolution does is to select sequences that allow proteins to exert their task more efficiently in the environment they live in while maintaining the same overall 3D architecture~\cite{Domingues2000,Karshikoff2015}.

Understanding which changes in the amino acid sequence can improve protein efficiency while preserving the biological function has both theoretical and practical implications.
Many works investigated the role of different amino acids in the protein structure, folding, stability and dynamics~\cite{Chakrabarty2016}. 
In this respect, methods based on graph theory approaches have contributed considerably to the understanding of protein structural flexibility, their hierarchy of structures and in the identification of key residues~\cite{Dokholyan2002,delSol2006, Amitai2004,pmid12188762, Aftabuddin2007}. All those findings demonstrated that a network-based analysis can be pivotal to shed light on the complex aspects relative to the organization of protein structures~\cite{Miotto2018}.
However, network approaches have often focused on a static description of the system while interesting properties, especially at the level of the single residue, are related to the dynamical behavior of the network~\cite{Yang2013}.

Theoretical epidemic modeling indeed is a typical approach to study the dynamical behavior of an interaction network, describing the evolution of a contagion process across a population~\cite{Kermack1927,Kermack1932}. 

In the last decades, epidemic models have seen applications in several fields \cite{Vespignani2011, Yang2016} thanks to the growth of network sciences.  From the spread of real diseases to the diffusion of news in social networks,  epidemic models give a measure about the diffusion of information within either the whole network or from a particular node to any other.

Here, we combine a graph-based schematization of proteins with an epidemic diffusion approach to study the overall stability and the capability to propagate perturbations (or information) in their complex 3D-structures~\cite{Castellano2009, Albert2002}.

In particular, our novel approach proved to be very efficient in characterizing protein thermal stability and in identifying functional sites of proteins, where trivial static network descriptors exhibit a lower efficiency.

\section*{Methods}

\begin{figure*}[!]
\centering
\includegraphics[width=\textwidth]{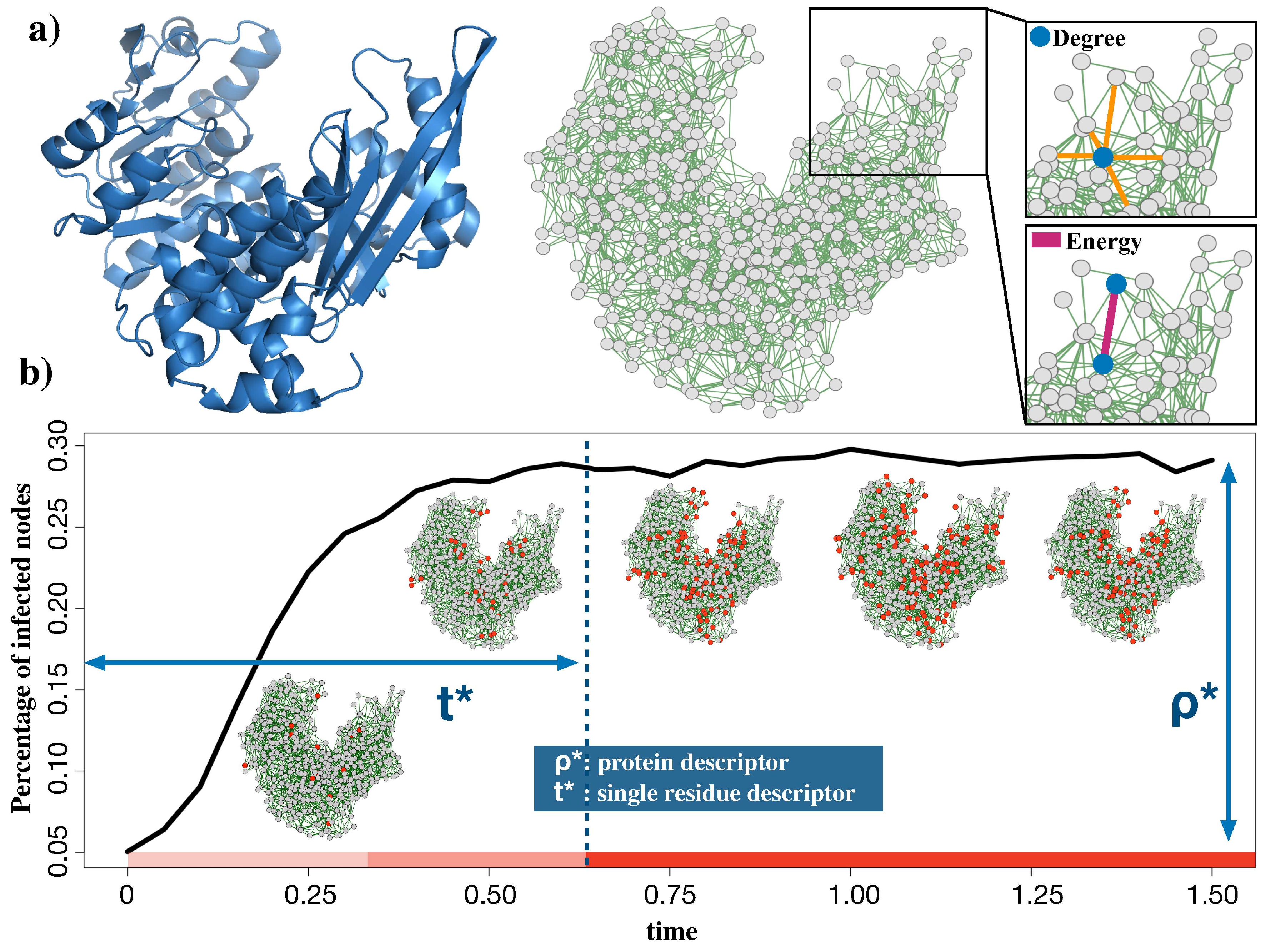}
\caption{\textbf{Scheme of the diffusion procedure.} \textbf{a)} Representation of the glutamate dehydrogenase (PDB id: 1HRD) protein  structure as ribbons (left) and as a Residue Interaction Network (RIN). 
 Protein residues are considered as nodes and the non-bonded energetic interactions between residues constitute the links between nodes. 
 \textbf{b)} Outcomes of an epidemic diffusion using interaction energy between residues and node degree as a proxy of infection and recover probability, respectively (as displayed in panel a). Two parameters can be defined:  the density of infected nodes at the stationary state, $\rho^\star$, and the time necessary to reach the equilibrium value, $t^\star$. The red nodes represent infected residues at a different time of the epidemic time evolution. 
 }
\label{fig:1}
\end{figure*}

\subsection{Datasets}
To investigate the capability of a diffusion protocol to grasp the essential feature of the protein structure and function, we defined four different datasets:  `Thermal dataset', `Enzyme dataset', `Allosteric dataset' and `HIV dataset'. Details regarding their collection are provided below.
\begin{itemize}
\item \textbf{Thermal dataset.} A set of 32 pairs of homologous proteins with different thermal properties was manually collected from literature~\cite{pmid16150969, pmid11161106, MozoVillaras2003, Sterner2001}. Experimentally determined structures were collected from the PDB~\cite{pmid25352545} and filtered according to method (x-ray diffraction), resolution (below 3 $\AA$), and percentage of missing residues (covering more than 95\% of to the Uniprot~\cite{pmid28150232} sequence). Proteins for which experimentally determined structures were only available in a bound state, i.e. in complex with either a ligand or an ion, were excluded. Further information is available in  Table 1 of the Supplementary Material). 
\item \textbf{Enzyme dataset.} It was composed grouping all the enzymes present among the proteins of the Thermal dataset. For each enzyme, we retrieved information about the residues forming the active site (see Table~1 in the Supplementary Material), from the Enzyme Portal of EBI~\cite{Alcntara2012}.

\item  \textbf{Allosteric dataset.} We collected from~\cite{Amor2016} proteins whose active and allosteric sites are both known. 

\item \textbf{HIV dataset.} It is composed of 2 apo (free) structures of the  HIV-1  e HIV-2 proteases together with 16 holo (bound) PDB structures (8 of HIV-1 and 8 of HIV-2) being in complex with different ligands is taken  from~\cite{triki2018exploration}.

\end{itemize}

All protein structures were minimized using the standard NAMD~\cite{pmid16222654} algorithm and the
CHARMM force field~\cite{pmid23146088} in vacuum. A 1 fs time step was used and structures were allowed to thermalize for 10000 time steps. This procedure aims at removing energetic clashes that may be present due to the crystallization procedure.

\begin{figure*}[!]
\centering
\includegraphics[width=\textwidth]{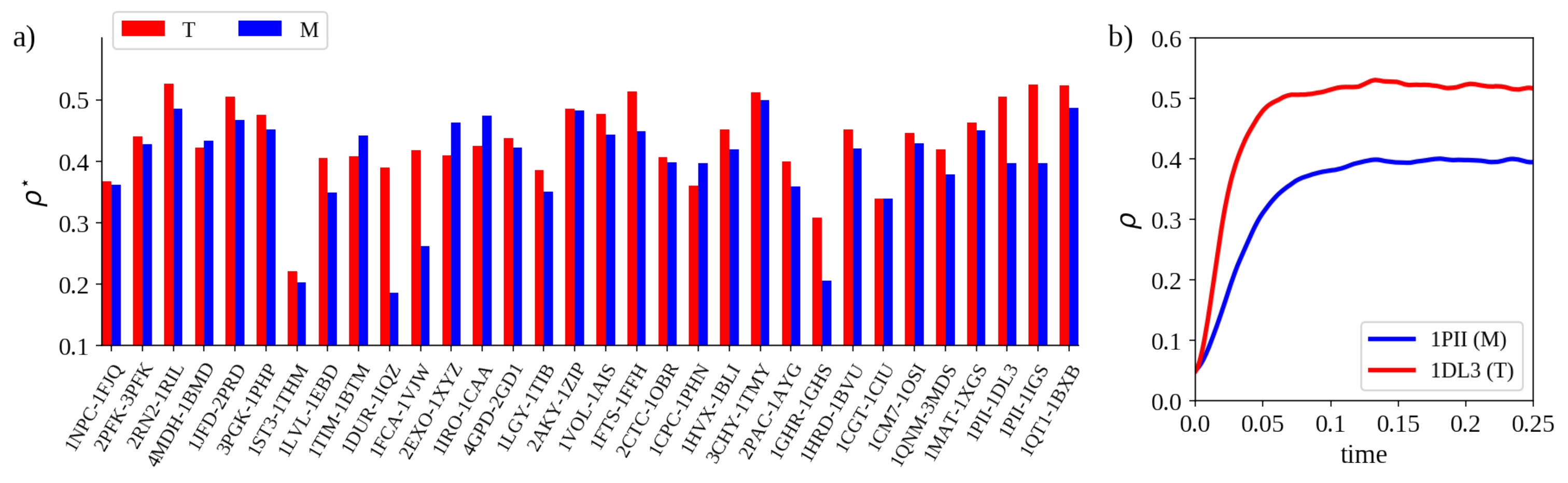}
\caption{ \textbf{The stationary density of infected nodes gives information on the thermal stability of the protein.} \textbf{a)} Mean density of infected nodes as a function of time for an explicative homologous couple. \textbf{b)} Bar-plot representation of the density of infected node at the stationary state for the 32 mesostable (blue) and thermostable (red) proteins of the Thermal dataset. 
}
\label{fig:2}
\end{figure*}

\subsection{Network representation}
Protein structures are represented as Residue Interaction Networks~\cite{pmid26216263} (RINs in short), where each node represents a single residue $aa_i$. The nearest atomic distance between a given pair of residues $aa_i$ and $aa_j$ is defined as $D_{ij}$. Two RIN nodes are linked together if $D_{ij} \leq 12~\AA$ \cite{pmid16222654, pmid23146088}. Furthermore links are weighted by the sum of two energetic terms: Coulomb (C) and Lennard-Jones (LJ) potentials. The C contribution between two atoms, $a_l$ and $a_m$, is calculated as:

\beq
\label{eq:C}
E_{lm}^C= \frac{1}{4\pi\epsilon_0}\frac{q_lq_m}{r_{lm}}
\eeq

where $q_l$ and $q_m$ are the partial charges for atoms $a_l$ and $a_m$, as obtained from the CHARMM force-field: $r_{lm}$ is the distance between the two atoms, and $\epsilon_0$ is the vacuum permittivity. The Lennard-Jones potential is instead given by:
\beq
\label{eq:LJ}
E_{lm}^{LJ} = \sqrt{\epsilon_l \epsilon_m}\left[ \left(\frac{R_{min}^l + R_{min}^m}{r_{lm}}\right)^{12} - 2\left(\frac{R_{min}^l + R_{min}^m}{r_{lm}}\right)^{6}\right]
\eeq
where $\epsilon_l$ and $\epsilon_m$ are the depths of the potential wells of atom l and m respectively, $R_{min}^l$  and $R_{min}^m$ are the distances at which the potentials reach their minima.
Therefore, the weight of the link connecting residues $aa_i$ and $aa_j$ is calculated by summing the contribution of the single atom pairs as:
\beq
\label{eq:etot}
E_{ij} = \left[ \sum_{l}^{N_i} \sum_{m}^{N_j} \left( E_{lm}^C + E_{lm}^{LJ} \right) \right]
\eeq

where $N_i$ and $N_j$ are the numbers of atoms of the i-th and j-th residue respectively.

\subsection{Diffusion model on the protein network}
 
Epidemic modeling describes the dynamical evolution of the contagion process within a population. An individual (or node) is said susceptible (S) when it is healthy but could contract the disease, infected (I) when the contagion is transmitted by an adjacent node and recovered (R) when it manages to recover from the disease. In principle, recovered individuals are immunized and hence they are safe from other infections for a certain time. 
To study the evolution of the density of infected individuals we have to define the basic processes that rule the transition of individuals  states, e.g.
\begin{equation}
\label{eq:rates}
\begin{cases}
\text{Susceptible to Infected} & (S \rightarrow I)\\
\text{Infected to Recovered} &(I \rightarrow R)\\
\text{Recovered to Susceptible}& (R \rightarrow S)\\
\end{cases}
\end{equation}

More in details, we must specify (i) the topology of the interaction network, i.e. which nodes directly interact with each other; (ii) the strength of the interaction which is linked to the transmission rate of the infection and (iii) the recovering rate (i.e. the probability, if present, of return healthy after having contracted the infection).   
Depending on the choices one makes for the set of transitions in Eq.~\ref{eq:rates}, different models and processes can be simulated.
A  detailed description of the most studied models in classical epidemiology is given in~[\cite{RevModPhys.87.925}.

In the present work, we simulated an epidemic diffusion over the protein RINs (see Figure~\ref{fig:1}). While we preserved the full topological information, we restrained to an SIS (susceptible-infected-susceptible) epidemic model, where each node (i.e. residue) once infected,  can transmit the infection to near neighbor nodes in the network.  
A residue can recover from the infection, returning to the susceptible state (meaning that the transition $R\rightarrow S$ is instantaneous). 
In this scenario, the probability of finding a node $i$ in infected state is given by:

\begin{equation}
\label{eq:pt}
p_{t+1}^i= (1-\delta^i)p_t^i + (1-p_t^i)\sum_{a=1}^N \beta_{ij}p^j_t    \end{equation}

where $\delta^i$ is the rate with which  node $i$ recovers from infection, while $\beta_{ij}$ represents the infection rate of node $i$ given that node $j$ is infected at time $t$~\cite{allen1994some}.  

Eq.~\ref{eq:pt} can not be solved analytically for complex topologies like the RIN ones but a numerical treatment is required.
In the Supplementary Material, we provide a short treatment of the mean-field approximation, where instead it is possible to analytically solve  Eq.~\ref{eq:pt} for different choices of the transitions in Eq.~\ref{eq:rates}.
In our case,  for each RIN node we identify the recovering rate, $\delta_i$, with the node degree. While the infection rate between node $i$ and $j$, $\beta_{ij}$ is given by the weight of the link ($\beta_{ij} = E_{ij}$ as given by Eq.~\ref{eq:etot}). 
Once defined the infection and recovering rates, we simulated the diffusion process into the 3D structure, starting from a specific set of residues or by picking an initially random set, and looking at the mean density of infected residues over time:

\begin{equation}
\rho(t) = \left<\frac{N_I(t)}{N_{tot}}\right>
\end{equation}

where $N_I(t)$ is the number of infected residue at time t, $N_{tot}$ is the total number of protein residues and we indicated with $<\cdot>$ the mean over the $M$ realizations of the diffusion process ( all results presented here, are obtained setting M=1000 in order to avoid large fluctuations that are unavoidable in a single realization).
It has been found that, depending on the connectivity matrix architecture and the sets of $\{\delta_{i}\}$ and $\{\beta_{ij}\}$ parameters, the system can exhibit different behaviors. As $t\rightarrow \infty$, the infection, starting from some nodes,  propagates in the whole network and reaches a stationary regime where a certain density $\rho^\star$ of nodes is constantly infected at each time, independently from the size and the identity of the initial set of infected nodes.
Intuitively, $\rho^\star = 0$ if the number of nodes that recover from the infection overcomes those that become infected. On the other hand, $\rho^\star = 1$ when the infection is too aggressive.
The nontrivial scenario ($0 < \rho^\star < 1$) is achieved when the network architecture and the parameters allow having a balance between the number of nodes that become infected and the ones that recover.
In the simulations, we defined  the transient time $t^\star$ as  the time after which $\rho(t^\star) = \rho^\star- \delta$, with $\delta \rightarrow 0$ (see Figure~\ref{fig:1}b); in other words $t^\star$ is the number of time steps needed by the epidemic to reach its stationary state.  

Statistical analysis was performed by using R package stats~\cite{rr}. In particular, clustering analysis performed on the HIV dataset was made using the HeatMap function, applying the Euclidean distance matrix (given by the `dist' function) and the `hclust' method for the clustering algorithm.  

\section*{Results}

\begin{figure*}[!]
\centering
\includegraphics[width=\textwidth]{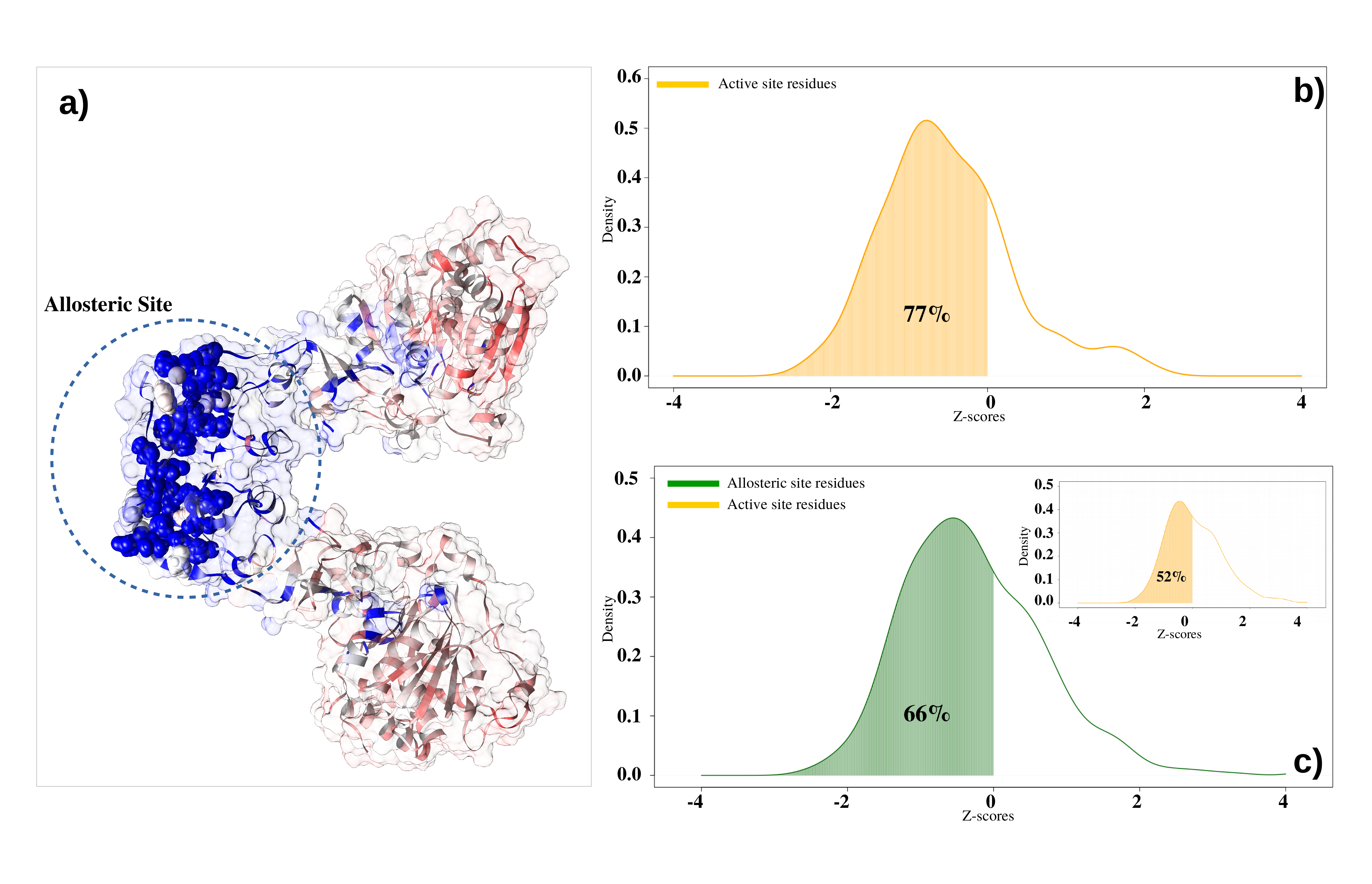}
\caption{\textbf{An epidemics starting from a functional site of protein spreads faster.} \textbf{a)}We report an example of a protein functional site mapped by using diffusion based approach. Ribbon representation of protein, identified by PDB code 1D09, was colored according $t^\star$, from red (highest values) to blue (lowest values). It is clear that allosteric site (residues reported as spheres) present lowest values of $t^\star$ parameter. \textbf{b)}  Distribution of the $z$-score values regarding the active site amino acids of the 24 enzymes in the Thermal dataset. \textbf{c)}    Distribution of the $z$-score values regarding the active sites and allosteric sites amino acids of the 20 proteins in the Allosteric dataset.}
\label{fig:3}
\end{figure*}

\subsection{Stationary epidemic behavior is a global measure of protein thermal stability}

Different thermal behaviors in homologous proteins have long been studied and several features have been identified as responsible for those differences (such as salt bridges, charged amino acids disposition, etc. ~\cite{Amadei_2017, Tartaglia_2007, VISHVESHWARA2002, Vijayabaskar2010, pmid25393107,pmid20159163, pmid18161956}).

These features are very well defined in network representation, both in terms of network topology (structure) and link weights (energy). 
Here we exploit our epidemic-diffusion algorithm to assess the capability of the network to reflect the protein thermostability.

In particular, we compared the stationary state density of infected nodes between all the couples of the Thermal dataset. For each protein, the diffusion was simulated, starting each time from a randomly selected set of infected residues. In particular, $5\%$ of the nodes were infected at $t=0$. 

In 84\% of cases (27 out of 32 comparisons), thermophilic proteins acquired a higher density of infected nodes with respect to their mesophilic counterparts, when epidemic diffusion reaches the equilibrium (Figure~\ref{fig:2}a). According to us this results reflects both the overall higher connectivity and the higher energy of the links in the thermophilic proteins compared to the mesophilic ones. Our diffusion-based approach is able to well capture this aspect, even better than the network analysis alone is able to do (see supplementary material). In Figure~\ref{fig:2}b, we reported an example of diffusion process results where the different steady states are very well visible (PDB id: 1PII-1DL3).

\subsection{Epidemic transient phase permits local characterization of protein structures}

After demonstrating that we can properly apply the epidemic diffusion approach exploring global features of a three-dimensional structure of a protein, we investigated our diffusion approach at a single residue level. i.e. we tested if residues that functionally need to have strong communication with the rest of the protein are characterized by peculiar diffusive properties. In this framework, one of the most important challenges in computational biology is the characterization of the active and allosteric sites in proteins. Since the substrate-binding has to be detected also far from the binding region through a cascade of residue-residue interactions~\cite{Amor2016}, we hypothesize that the diffusive approach could be a perfect approach to capture this aspect.

So we investigated, in particular, the transient phase of the epidemic, i.e. the number of time steps necessary to reach the equilibrium ($t^\star $). The time $t^\star $ varies according to some features of the infected initial nodes. In particular, if the epidemic starts from energetically interconnected residues, it is very likely that the stationary state will be obtained in a shorter time when compared with the other residues.

We simulated an epidemic originating from every single residue of all proteins in the Enzyme dataset. In particular, since epidemic originating from a single node usually are characterized by a fast extinction, for each residue, we selected also its two closest neighbors in sequence as a contagion starting point.

We normalized results over each protein size by using the $z$-score, in order to make the comparison between proteins with sequences of different length possible.

The $z$-score of the ${i}$-th residue was defined as: 

\beq
\label{eq:zscore}
z_i = \frac{t^\star_i- \overline{t^\star}}{\sqrt{\overline{(t^\star)^2} - \left(\overline{t^\star}\right)^2}}
\eeq

where the over-bar represents the mean of $t^\star$ over all the amino acids in the analyzed protein.

Charged residues exposed on the protein surface and core residues are obviously very fast in propagating the infection, because of the high energy interactions the charged residues are involved in and because of the high number of contacts the core residues have. We preliminary confirmed this (see supplementary material) as shown in Suppl. Figure 1. We also correlated protein secondary structure location (as calculated by STRIDE~\cite{Frishman1995}) of each residue with its $t^\star$ because we can suppose that, on average, residues belonging to secondary structures are assembled in a dense part of the interaction network. As expected, Beta Strand and Alpha Helix components are characterized by lower $t^\star$ (see supplementary material).

We then proceeded to apply the diffusion protocol in order to analyze $t^\star $ values for 11 pairs of (thermostable-mesostable) enzymes in the Enzyme dataset for which we know residues forming the active site. The comparison between the $t^\star$ values of the active site with that of the other residues clearly shows that the former is characterized by a statistically significant lower $t^\star$. The `functional information' the protein receives after ligand binding into the active site needs to be fastly communicated to the whole protein and our diffusive method is able to well characterize this important biological aspect. We show in Figure~\ref{fig:3}b the distribution of $z$-scores belonging to the active site residues. The $77\%$ of residues belonging to active sites presents a $z$-score lower than 0,   represented by the orange area under the curve. This means stronger connectivity of active site residues with the whole protein than the average value of all other residues.   

Then we have considered the twenty allosteric proteins with known active and allosteric site residues (see  Allosteric dataset in Methods). Even in this case, we applied our epidemic protocol in order to evaluate the number of time steps necessary to reach the equilibrium.
As shown in Figure~\ref{fig:3}c, $66\%$ of allosteric residues shows a $z$-score value lower than zero, demonstrating that allosteric residues are faster than average residues in propagating information inside the protein network due to their biological functional role in the 3D structure. Interestingly, the active sites of these proteins are not characterized by peculiar diffusive characteristics because just $52\%$ active sites reach $z$-score values lower than 0. The reason for  this behavior, different from what observed before, can be due to their different binding `state': in the Thermal dataset the proteins were in the apo form, while in the Allosteric dataset the proteins are in the holo form, with the ligand occupying the active site and diffusive approach seems to be very sensitive to these two states.

\subsection{HIV-1 and HIV-2 Proteases can be discriminated by their epidemic diffusion profile}

Finally, we used epidemic diffusion time analysis in order to discriminate HIV-1 and HIV-2 proteases. Despite the structural similarities, HIV-1 and HIV-2 proteases show dramatic disparities in susceptibility to HIV-1 protease inhibitors \cite{triki2018exploration}. Each of the 18 HIV proteases (HIV dataset) has been represented by a diffusion time profile (e.g. the concatenation of single residue $t^\star$)  and all profiles have been easily compared since all protein sequences have the same length.

\begin{figure*}[!]
\centering
\includegraphics[width=\textwidth]{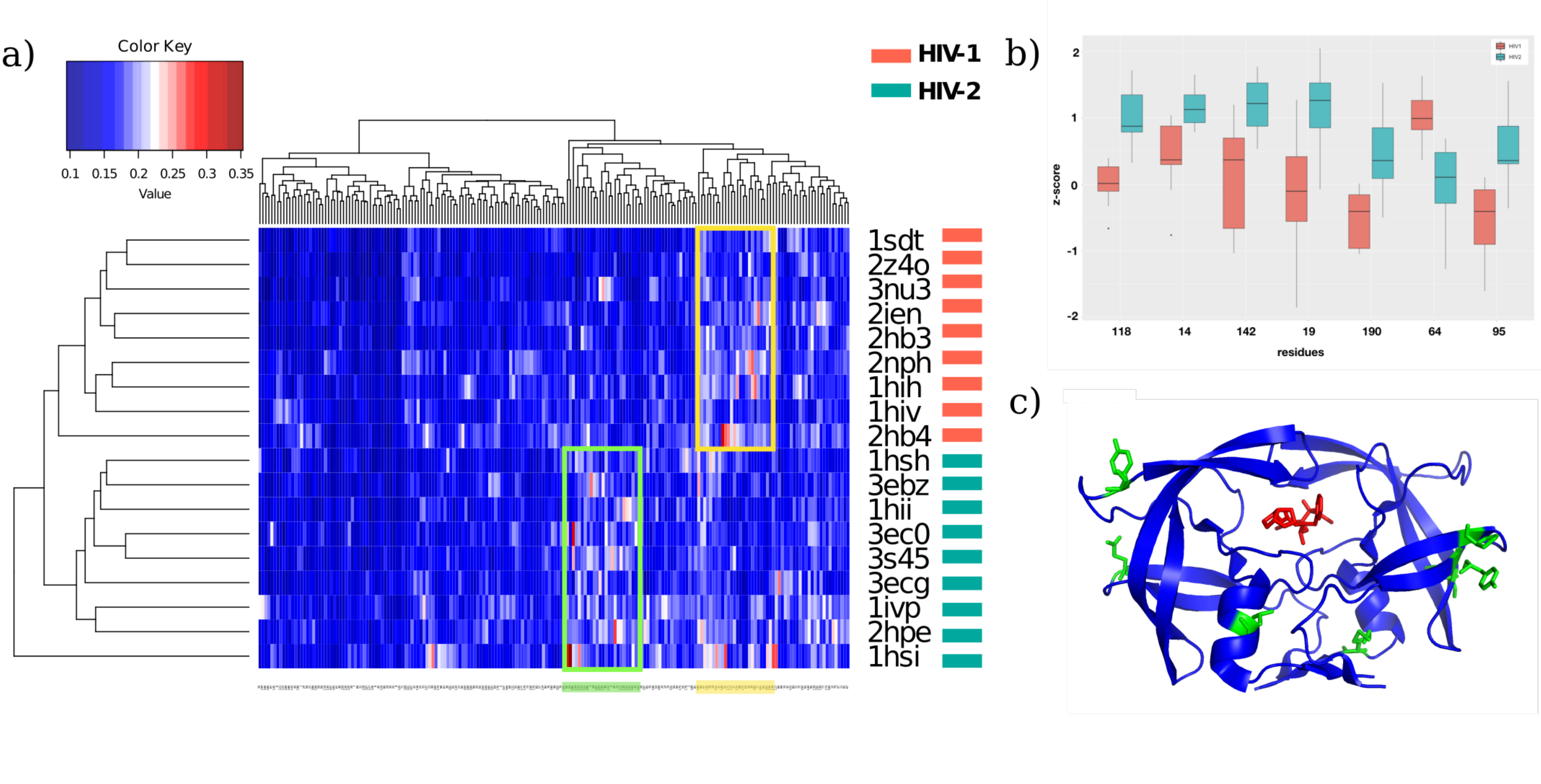}
\caption{\textbf{The epidemic diffusion profile discriminates different HIV proteases.}  \textbf{a)} Heatmap representation of the clustering analysis performed on the $z_i$ scores of the proteases of the HIV dataset. \textbf{b)} Boxplot of the  distributions of the $z_i$ scores of the 7 most different HIV residues. \textbf{c)} Representation of the HIV protease (PDB id: 3ECG). Green sticks highlight the seven residues most responsible for the difference between HIV-1 and HIV-2 protease sets.}
\label{fig:4}
\end{figure*}

The heatmap reported in Figure~\ref{fig:4}a shows clustering analysis results performed on residues and proteins of the HIV dataset. All of the proteins are correctly identified as HIV-1 or HIV-2 protease demonstrating the possibility to use an epidemic diffusion approach in order to evaluate functional differences related to three-dimensional protein structures. 

We then explored key residues responsible for discrimination of the two groups. For each residue of HIV-1 and HIV-2 proteases, we compared their $z_i$ scores distributions with a t-test. The 39 residues showing the most significant difference, when we set p-value threshold to 0.05  are: 14,  19,  22,  23,  40,  41,  43,  56,  61,  62,  64,  70,  72,  73,  84,  95, 103, 108, 114, 115, 116, 118, 120, 134, 136, 140, 142, 155, 160, 163, 165, 167, 168, 170, 179, 190, 191, 192, 193.

Notably, lowering the p-value threshold to 0.005, we identified a subgroup of seven residues, i.e. 14, 19, 64, 95,1 18, 142, 190, shown in Figure~\ref{fig:4}b. 
As we said before, although HIV-1 and HIV-2 proteases share a great deal of structural similarity the reasons for intrinsic protease inhibitor resistance in HIV-2 are not known. Very interestingly, the subgroup of seven residues we identified occur at sites distant from the active site (see Figure~\ref{fig:4}c) and two out of seven residues are also identical in the two proteases. This leads us to present a hypothesis that perhaps our epidemic approach could have been captured subtle structural changes, imparted by a limited number of residues, causing dramatic functional differences between homologous proteins (HIV-1 and HIV-2 proteases). That is the seven amino acids outside the HIV-2 protease active site may cause subtle changes in conformation and in long-range effects compared to HIV-1, which might impact protease inhibitors binding affinity.

\section{Discussion}

Proteins are complex systems where evolution must be very proficient in tuning parameter (e.g. selecting mutations) to obtained more fitted proteins with respect to some features while maintaining the  protein functional. For instance, optimizing enzymes to be more efficient at high temperatures (i.e. increasing their thermal stability) must not reduce enzyme flexibility and the ability to change configurations.

Graph theory-based methods  represent a powerful approach to investigate protein topological and energetic properties. However, we could consider it a static view of the protein structures that does not allow us to describe their complexity in a complete way. To overcome this limitation several aspects of proteins were investigated through dynamical approaches, like molecular dynamics or perturbation-response approaches, which take into account the dynamical properties. A problem connected with these approaches is that they are typically characterized by a high computational cost.

In this work, we explored the possibility to adopt an epidemic diffusion-based method as an efficient way to study functional aspects (both local and global) of proteins strictly connected with their three-dimensional structural organization. The new idea we introduced with this approach was to investigate the dynamic properties of the interactions network of a protein structure by using epidemic diffusion-based algorithms, preserving in this way both the topology and the energy properties of the interactions. The most striking advantage of this method is that it is not very computationally expensive allowing for a fast exploration of complex problems related to protein function (diffusion on an average protein requires few minutes on a standard personal computer). 
Starting from the RIN formalism~\cite{Sengupta2012, Bde2007}, we studied the diffusion of a fictitious epidemic inside the protein structure represented as a network using energies and node degrees as proxies of infection and recovery rates. 

A large number of mathematical models have been formulated to study the spread of infectious diseases, but most of these are just variants of Kermack and McKendrick epidemic model~\cite{Kermack1927,Kermack1932}. 
Reproducing different aspects of the spread of real diseases, all models ultimately provide a measure of the information diffusion throughout the entire network.


Simulations of diffusion processes were performed considering typical network parameters for calculating the probability of transmission
of infection (proportional to the link energy) and the probability of each node of returning susceptible (proportional to node degree). 

From diffusion simulations, two descriptors were defined, one ($\rho^\star$) providing global information and the other ($t^\star$) local one.
In particular, a residue-specific descriptor is of fundamental importance because the identification of functional key residues in a protein structure is
a useful aspect for protein design in many open biological questions. 

Considering the stationary phase, the mean of the percentage of infected nodes is constant over the steps balancing the rate of infection and recovery. The
The value of the stationary percentage of infected nodes is a very compact way to quantify the global properties of the entire protein related to residue-residue energetic interactions.
A protein characterized by strong interconnectivity will have a very strong energetic coupling between its residues showing,  at the equilibrium, a higher number of infected nodes.




Given an overall fold, the arrangement of side chains organizes the inter-molecular interaction to better resist the thermal noise. Therefore, we test the sensibility of this formalism applying it on a well-defined set of homologous protein pairs,  one protein from a mesostable organism, the other one from a thermostable one. 

We found that thermophilic proteins have a significantly higher percentage of infected residues than homologous mesophilic counterparts, meaning that thermophilic proteins organize their network of interactions in order to promote infection. We could, therefore, conclude that thermophilic proteins have, on average, a higher level of interconnectivity than mesophilic proteins.

Another important aspect we explored in this work was the local properties of proteins that often are generated by long-range effects. In this case, the problem was studied by taking into account the transient phase of the diffusion simulation, which is composed of steps between the initial infection and the stationary state. 

The number of steps necessary in order to reach the stationary phase, $t^\star $, is depending on the choice of the starting infected nodes, expressing their centrality in the energy network. 
This local characterization can be utilized in order to identify which kind of residues (or domains) are more central in a protein, in terms of their connection with the rest of the protein. We clearly demonstrated that both residues belonging to enzyme allosteric and active sites typically reach the state of equilibrium with a number of steps smaller than any other residue.

We also investigated the local property of each residue of two HIV-1 and HIV-2 proteases. The method showed its perfect ability to separate the two classes of proteases, in terms of transient phase, elucidating non-trivial differences (HIV-1 and HIV-2 protease share $50\%$ of sequence identity) by analyzing the dynamic properties of residues represented as a network. The epidemic approach was also able to select seven residues responsible for the discrimination of the two groups, which might impact protease inhibitors binding affinity helping to understand the key differences between HIV-1 and HIV-2 infections.
We believe that the study of the dynamical aspect of the protein structure network is, in general, a promising direction for the future. We intend to investigate how far our results can be generalized to other types of protein functional elements, other types of proteins like membrane proteins and hopefully to other kinds of macromolecules like nucleic acids.





\bibliography{biblioThermo}

\begin{thebibliography}{47}
\expandafter\ifx\csname natexlab\endcsname\relax\def\natexlab#1{#1}\fi
\expandafter\ifx\csname bibnamefont\endcsname\relax
  \def\bibnamefont#1{#1}\fi
\expandafter\ifx\csname bibfnamefont\endcsname\relax
  \def\bibfnamefont#1{#1}\fi
\expandafter\ifx\csname citenamefont\endcsname\relax
  \def\citenamefont#1{#1}\fi
\expandafter\ifx\csname url\endcsname\relax
  \def\url#1{\texttt{#1}}\fi
\expandafter\ifx\csname urlprefix\endcsname\relax\def\urlprefix{URL }\fi
\providecommand{\bibinfo}[2]{#2}
\providecommand{\eprint}[2][]{\url{#2}}

\bibitem[{\citenamefont{Mannige}(2014)}]{Mannige2014}
\bibinfo{author}{\bibfnamefont{R.}~\bibnamefont{Mannige}},
  \bibinfo{journal}{Proteomes} \textbf{\bibinfo{volume}{2}},
  \bibinfo{pages}{128} (\bibinfo{year}{2014}).

\bibitem[{\citenamefont{Chothia et~al.}(1997)\citenamefont{Chothia, Hubbard,
  Brenner, Barns, and Murzin}}]{Chothia1997}
\bibinfo{author}{\bibfnamefont{C.}~\bibnamefont{Chothia}},
  \bibinfo{author}{\bibfnamefont{T.}~\bibnamefont{Hubbard}},
  \bibinfo{author}{\bibfnamefont{S.}~\bibnamefont{Brenner}},
  \bibinfo{author}{\bibfnamefont{H.}~\bibnamefont{Barns}}, \bibnamefont{and}
  \bibinfo{author}{\bibfnamefont{A.}~\bibnamefont{Murzin}},
  \bibinfo{journal}{Annual Review of Biophysics and Biomolecular Structure}
  \textbf{\bibinfo{volume}{26}}, \bibinfo{pages}{597} (\bibinfo{year}{1997}).

\bibitem[{\citenamefont{Dill et~al.}(2008)\citenamefont{Dill, Ozkan, Shell, and
  Weikl}}]{Dill2008}
\bibinfo{author}{\bibfnamefont{K.~A.} \bibnamefont{Dill}},
  \bibinfo{author}{\bibfnamefont{S.~B.} \bibnamefont{Ozkan}},
  \bibinfo{author}{\bibfnamefont{M.~S.} \bibnamefont{Shell}}, \bibnamefont{and}
  \bibinfo{author}{\bibfnamefont{T.~R.} \bibnamefont{Weikl}},
  \bibinfo{journal}{Annual Review of Biophysics} \textbf{\bibinfo{volume}{37}},
  \bibinfo{pages}{289} (\bibinfo{year}{2008}).

\bibitem[{\citenamefont{Lesk and Chothia}(1980)}]{Lesk1980}
\bibinfo{author}{\bibfnamefont{A.~M.} \bibnamefont{Lesk}} \bibnamefont{and}
  \bibinfo{author}{\bibfnamefont{C.}~\bibnamefont{Chothia}},
  \bibinfo{journal}{Journal of Molecular Biology}
  \textbf{\bibinfo{volume}{136}}, \bibinfo{pages}{225} (\bibinfo{year}{1980}).

\bibitem[{\citenamefont{Ofran and Margalit}(2006)}]{Ofran2006}
\bibinfo{author}{\bibfnamefont{Y.}~\bibnamefont{Ofran}} \bibnamefont{and}
  \bibinfo{author}{\bibfnamefont{H.}~\bibnamefont{Margalit}},
  \bibinfo{journal}{Proteins: Structure, Function, and Bioinformatics}
  \textbf{\bibinfo{volume}{64}}, \bibinfo{pages}{275} (\bibinfo{year}{2006}).

\bibitem[{\citenamefont{Deb{\`{e}}s et~al.}(2013)\citenamefont{Deb{\`{e}}s,
  Wang, Caetano-Anoll{\'{e}}s, and Gr\"{a}ter}}]{Debs2013}
\bibinfo{author}{\bibfnamefont{C.}~\bibnamefont{Deb{\`{e}}s}},
  \bibinfo{author}{\bibfnamefont{M.}~\bibnamefont{Wang}},
  \bibinfo{author}{\bibfnamefont{G.}~\bibnamefont{Caetano-Anoll{\'{e}}s}},
  \bibnamefont{and}
  \bibinfo{author}{\bibfnamefont{F.}~\bibnamefont{Gr\"{a}ter}},
  \bibinfo{journal}{{PLoS} Computational Biology} \textbf{\bibinfo{volume}{9}},
  \bibinfo{pages}{e1002861} (\bibinfo{year}{2013}).

\bibitem[{\citenamefont{Domingues et~al.}(2000)\citenamefont{Domingues,
  Koppensteiner, and Sippl}}]{Domingues2000}
\bibinfo{author}{\bibfnamefont{F.~S.} \bibnamefont{Domingues}},
  \bibinfo{author}{\bibfnamefont{W.~A.} \bibnamefont{Koppensteiner}},
  \bibnamefont{and} \bibinfo{author}{\bibfnamefont{M.~J.} \bibnamefont{Sippl}},
  \bibinfo{journal}{{FEBS} Letters} \textbf{\bibinfo{volume}{476}},
  \bibinfo{pages}{98} (\bibinfo{year}{2000}).

\bibitem[{\citenamefont{Karshikoff et~al.}(2015)\citenamefont{Karshikoff,
  Nilsson, and Ladenstein}}]{Karshikoff2015}
\bibinfo{author}{\bibfnamefont{A.}~\bibnamefont{Karshikoff}},
  \bibinfo{author}{\bibfnamefont{L.}~\bibnamefont{Nilsson}}, \bibnamefont{and}
  \bibinfo{author}{\bibfnamefont{R.}~\bibnamefont{Ladenstein}},
  \bibinfo{journal}{{FEBS} Journal} \textbf{\bibinfo{volume}{282}},
  \bibinfo{pages}{3899} (\bibinfo{year}{2015}).

\bibitem[{\citenamefont{Chakrabarty and Parekh}(2016)}]{Chakrabarty2016}
\bibinfo{author}{\bibfnamefont{B.}~\bibnamefont{Chakrabarty}} \bibnamefont{and}
  \bibinfo{author}{\bibfnamefont{N.}~\bibnamefont{Parekh}},
  \bibinfo{journal}{Nucleic Acids Research} \textbf{\bibinfo{volume}{44}},
  \bibinfo{pages}{W375} (\bibinfo{year}{2016}).

\bibitem[{\citenamefont{Dokholyan et~al.}(2002)\citenamefont{Dokholyan, Li,
  Ding, and Shakhnovich}}]{Dokholyan2002}
\bibinfo{author}{\bibfnamefont{N.~V.} \bibnamefont{Dokholyan}},
  \bibinfo{author}{\bibfnamefont{L.}~\bibnamefont{Li}},
  \bibinfo{author}{\bibfnamefont{F.}~\bibnamefont{Ding}}, \bibnamefont{and}
  \bibinfo{author}{\bibfnamefont{E.~I.} \bibnamefont{Shakhnovich}},
  \bibinfo{journal}{Proceedings of the National Academy of Sciences}
  \textbf{\bibinfo{volume}{99}}, \bibinfo{pages}{8637} (\bibinfo{year}{2002}).

\bibitem[{\citenamefont{del Sol et~al.}(2006)\citenamefont{del Sol, Fujihashi,
  Amoros, and Nussinov}}]{delSol2006}
\bibinfo{author}{\bibfnamefont{A.}~\bibnamefont{del Sol}},
  \bibinfo{author}{\bibfnamefont{H.}~\bibnamefont{Fujihashi}},
  \bibinfo{author}{\bibfnamefont{D.}~\bibnamefont{Amoros}}, \bibnamefont{and}
  \bibinfo{author}{\bibfnamefont{R.}~\bibnamefont{Nussinov}},
  \bibinfo{journal}{Molecular Systems Biology} \textbf{\bibinfo{volume}{2}}
  (\bibinfo{year}{2006}).

\bibitem[{\citenamefont{Amitai et~al.}(2004)\citenamefont{Amitai, Shemesh,
  Sitbon, Shklar, Netanely, Venger, and Pietrokovski}}]{Amitai2004}
\bibinfo{author}{\bibfnamefont{G.}~\bibnamefont{Amitai}},
  \bibinfo{author}{\bibfnamefont{A.}~\bibnamefont{Shemesh}},
  \bibinfo{author}{\bibfnamefont{E.}~\bibnamefont{Sitbon}},
  \bibinfo{author}{\bibfnamefont{M.}~\bibnamefont{Shklar}},
  \bibinfo{author}{\bibfnamefont{D.}~\bibnamefont{Netanely}},
  \bibinfo{author}{\bibfnamefont{I.}~\bibnamefont{Venger}}, \bibnamefont{and}
  \bibinfo{author}{\bibfnamefont{S.}~\bibnamefont{Pietrokovski}},
  \bibinfo{journal}{Journal of Molecular Biology}
  \textbf{\bibinfo{volume}{344}}, \bibinfo{pages}{1135} (\bibinfo{year}{2004}).

\bibitem[{\citenamefont{Vendruscolo et~al.}(2002)\citenamefont{Vendruscolo,
  Dokholyan, Paci, and Karplus}}]{pmid12188762}
\bibinfo{author}{\bibfnamefont{M.}~\bibnamefont{Vendruscolo}},
  \bibinfo{author}{\bibfnamefont{N.~V.} \bibnamefont{Dokholyan}},
  \bibinfo{author}{\bibfnamefont{E.}~\bibnamefont{Paci}}, \bibnamefont{and}
  \bibinfo{author}{\bibfnamefont{M.}~\bibnamefont{Karplus}},
  \bibinfo{journal}{Phys Rev E Stat Nonlin Soft Matter Phys}
  \textbf{\bibinfo{volume}{65}}, \bibinfo{pages}{061910}
  (\bibinfo{year}{2002}).

\bibitem[{\citenamefont{Aftabuddin and Kundu}(2007)}]{Aftabuddin2007}
\bibinfo{author}{\bibfnamefont{M.}~\bibnamefont{Aftabuddin}} \bibnamefont{and}
  \bibinfo{author}{\bibfnamefont{S.}~\bibnamefont{Kundu}},
  \bibinfo{journal}{Biophysical Journal} \textbf{\bibinfo{volume}{93}},
  \bibinfo{pages}{225} (\bibinfo{year}{2007}).

\bibitem[{\citenamefont{Miotto et~al.}(2018)\citenamefont{Miotto, Olimpieri,
  Rienzo, Ambrosetti, Corsi, Lepore, Tartaglia, and Milanetti}}]{Miotto2018}
\bibinfo{author}{\bibfnamefont{M.}~\bibnamefont{Miotto}},
  \bibinfo{author}{\bibfnamefont{P.~P.} \bibnamefont{Olimpieri}},
  \bibinfo{author}{\bibfnamefont{L.~D.} \bibnamefont{Rienzo}},
  \bibinfo{author}{\bibfnamefont{F.}~\bibnamefont{Ambrosetti}},
  \bibinfo{author}{\bibfnamefont{P.}~\bibnamefont{Corsi}},
  \bibinfo{author}{\bibfnamefont{R.}~\bibnamefont{Lepore}},
  \bibinfo{author}{\bibfnamefont{G.~G.} \bibnamefont{Tartaglia}},
  \bibnamefont{and}
  \bibinfo{author}{\bibfnamefont{E.}~\bibnamefont{Milanetti}},
  \bibinfo{journal}{Bioinformatics}  (\bibinfo{year}{2018}).

\bibitem[{\citenamefont{Yang et~al.}(2013)\citenamefont{Yang, Sang, Tao, Fu,
  Zhang, Xie, and Liu}}]{Yang2013}
\bibinfo{author}{\bibfnamefont{L.-Q.} \bibnamefont{Yang}},
  \bibinfo{author}{\bibfnamefont{P.}~\bibnamefont{Sang}},
  \bibinfo{author}{\bibfnamefont{Y.}~\bibnamefont{Tao}},
  \bibinfo{author}{\bibfnamefont{Y.-X.} \bibnamefont{Fu}},
  \bibinfo{author}{\bibfnamefont{K.-Q.} \bibnamefont{Zhang}},
  \bibinfo{author}{\bibfnamefont{Y.-H.} \bibnamefont{Xie}}, \bibnamefont{and}
  \bibinfo{author}{\bibfnamefont{S.-Q.} \bibnamefont{Liu}},
  \bibinfo{journal}{Journal of Biomolecular Structure and Dynamics}
  \textbf{\bibinfo{volume}{32}}, \bibinfo{pages}{372} (\bibinfo{year}{2013}).

\bibitem[{\citenamefont{Kermack and McKendrick}(1927)}]{Kermack1927}
\bibinfo{author}{\bibfnamefont{W.~O.} \bibnamefont{Kermack}} \bibnamefont{and}
  \bibinfo{author}{\bibfnamefont{A.~G.} \bibnamefont{McKendrick}},
  \bibinfo{journal}{Proceedings of the Royal Society A: Mathematical, Physical
  and Engineering Sciences} \textbf{\bibinfo{volume}{115}},
  \bibinfo{pages}{700} (\bibinfo{year}{1927}).

\bibitem[{\citenamefont{Kermack and McKendrick}(1932)}]{Kermack1932}
\bibinfo{author}{\bibfnamefont{W.~O.} \bibnamefont{Kermack}} \bibnamefont{and}
  \bibinfo{author}{\bibfnamefont{A.~G.} \bibnamefont{McKendrick}},
  \bibinfo{journal}{Proceedings of the Royal Society A: Mathematical, Physical
  and Engineering Sciences} \textbf{\bibinfo{volume}{138}}, \bibinfo{pages}{55}
  (\bibinfo{year}{1932}).

\bibitem[{\citenamefont{Vespignani}(2011)}]{Vespignani2011}
\bibinfo{author}{\bibfnamefont{A.}~\bibnamefont{Vespignani}},
  \bibinfo{journal}{Nature Physics} \textbf{\bibinfo{volume}{8}},
  \bibinfo{pages}{32} (\bibinfo{year}{2011}).

\bibitem[{\citenamefont{Yang and Wang}(2016)}]{Yang2016}
\bibinfo{author}{\bibfnamefont{H.-X.} \bibnamefont{Yang}} \bibnamefont{and}
  \bibinfo{author}{\bibfnamefont{B.-H.} \bibnamefont{Wang}},
  \bibinfo{journal}{Physica A: Statistical Mechanics and its Applications}
  \textbf{\bibinfo{volume}{443}}, \bibinfo{pages}{86} (\bibinfo{year}{2016}).

\bibitem[{\citenamefont{Castellano et~al.}(2009)\citenamefont{Castellano,
  Fortunato, and Loreto}}]{Castellano2009}
\bibinfo{author}{\bibfnamefont{C.}~\bibnamefont{Castellano}},
  \bibinfo{author}{\bibfnamefont{S.}~\bibnamefont{Fortunato}},
  \bibnamefont{and} \bibinfo{author}{\bibfnamefont{V.}~\bibnamefont{Loreto}},
  \bibinfo{journal}{Reviews of Modern Physics} \textbf{\bibinfo{volume}{81}},
  \bibinfo{pages}{591} (\bibinfo{year}{2009}).

\bibitem[{\citenamefont{Albert and Barab{\'{a}}si}(2002)}]{Albert2002}
\bibinfo{author}{\bibfnamefont{R.}~\bibnamefont{Albert}} \bibnamefont{and}
  \bibinfo{author}{\bibfnamefont{A.-L.} \bibnamefont{Barab{\'{a}}si}},
  \bibinfo{journal}{Reviews of Modern Physics} \textbf{\bibinfo{volume}{74}},
  \bibinfo{pages}{47} (\bibinfo{year}{2002}).

\bibitem[{\citenamefont{Brinda and Vishveshwara}(2005)}]{pmid16150969}
\bibinfo{author}{\bibfnamefont{K.~V.} \bibnamefont{Brinda}} \bibnamefont{and}
  \bibinfo{author}{\bibfnamefont{S.}~\bibnamefont{Vishveshwara}},
  \bibinfo{journal}{Biophys. J.} \textbf{\bibinfo{volume}{89}},
  \bibinfo{pages}{4159} (\bibinfo{year}{2005}).

\bibitem[{\citenamefont{Kannan and Vishveshwara}(2000)}]{pmid11161106}
\bibinfo{author}{\bibfnamefont{N.}~\bibnamefont{Kannan}} \bibnamefont{and}
  \bibinfo{author}{\bibfnamefont{S.}~\bibnamefont{Vishveshwara}},
  \bibinfo{journal}{Protein Eng.} \textbf{\bibinfo{volume}{13}},
  \bibinfo{pages}{753} (\bibinfo{year}{2000}).

\bibitem[{\citenamefont{Mozo-Villarias
  et~al.}(2003)\citenamefont{Mozo-Villarias, Cedano, and
  Querol}}]{MozoVillaras2003}
\bibinfo{author}{\bibfnamefont{A.}~\bibnamefont{Mozo-Villarias}},
  \bibinfo{author}{\bibfnamefont{J.}~\bibnamefont{Cedano}}, \bibnamefont{and}
  \bibinfo{author}{\bibfnamefont{E.}~\bibnamefont{Querol}},
  \bibinfo{journal}{Protein Engineering, Design and Selection}
  \textbf{\bibinfo{volume}{16}}, \bibinfo{pages}{279} (\bibinfo{year}{2003}).

\bibitem[{\citenamefont{hard Sterner and Liebl}(2001)}]{Sterner2001}
\bibinfo{author}{\bibfnamefont{R.}~\bibnamefont{hard Sterner}}
  \bibnamefont{and} \bibinfo{author}{\bibfnamefont{W.}~\bibnamefont{Liebl}},
  \bibinfo{journal}{Critical Reviews in Biochemistry and Molecular Biology}
  \textbf{\bibinfo{volume}{36}}, \bibinfo{pages}{39} (\bibinfo{year}{2001}).

\bibitem[{\citenamefont{Touw et~al.}(2015)\citenamefont{Touw, Baakman, Black,
  te~Beek, Krieger, Joosten, and Vriend}}]{pmid25352545}
\bibinfo{author}{\bibfnamefont{W.~G.} \bibnamefont{Touw}},
  \bibinfo{author}{\bibfnamefont{C.}~\bibnamefont{Baakman}},
  \bibinfo{author}{\bibfnamefont{J.}~\bibnamefont{Black}},
  \bibinfo{author}{\bibfnamefont{T.~A.} \bibnamefont{te~Beek}},
  \bibinfo{author}{\bibfnamefont{E.}~\bibnamefont{Krieger}},
  \bibinfo{author}{\bibfnamefont{R.~P.} \bibnamefont{Joosten}},
  \bibnamefont{and} \bibinfo{author}{\bibfnamefont{G.}~\bibnamefont{Vriend}},
  \bibinfo{journal}{Nucleic Acids Res.} \textbf{\bibinfo{volume}{43}},
  \bibinfo{pages}{D364} (\bibinfo{year}{2015}).

\bibitem[{\citenamefont{Pundir et~al.}(2017)\citenamefont{Pundir, Martin, and
  O'Donovan}}]{pmid28150232}
\bibinfo{author}{\bibfnamefont{S.}~\bibnamefont{Pundir}},
  \bibinfo{author}{\bibfnamefont{M.~J.} \bibnamefont{Martin}},
  \bibnamefont{and}
  \bibinfo{author}{\bibfnamefont{C.}~\bibnamefont{O'Donovan}},
  \bibinfo{journal}{Methods Mol. Biol.} \textbf{\bibinfo{volume}{1558}},
  \bibinfo{pages}{41} (\bibinfo{year}{2017}).

\bibitem[{\citenamefont{Alc{\'{a}}ntara
  et~al.}(2012)\citenamefont{Alc{\'{a}}ntara, Onwubiko, Cao, de~Matos, Cham,
  Jacobsen, Holliday, Fischer, Rahman, Jassal et~al.}}]{Alcntara2012}
\bibinfo{author}{\bibfnamefont{R.}~\bibnamefont{Alc{\'{a}}ntara}},
  \bibinfo{author}{\bibfnamefont{J.}~\bibnamefont{Onwubiko}},
  \bibinfo{author}{\bibfnamefont{H.}~\bibnamefont{Cao}},
  \bibinfo{author}{\bibfnamefont{P.}~\bibnamefont{de~Matos}},
  \bibinfo{author}{\bibfnamefont{J.~A.} \bibnamefont{Cham}},
  \bibinfo{author}{\bibfnamefont{J.}~\bibnamefont{Jacobsen}},
  \bibinfo{author}{\bibfnamefont{G.~L.} \bibnamefont{Holliday}},
  \bibinfo{author}{\bibfnamefont{J.~D.} \bibnamefont{Fischer}},
  \bibinfo{author}{\bibfnamefont{S.~A.} \bibnamefont{Rahman}},
  \bibinfo{author}{\bibfnamefont{B.}~\bibnamefont{Jassal}},
  \bibnamefont{et~al.}, \bibinfo{journal}{Nucleic Acids Research}
  \textbf{\bibinfo{volume}{41}}, \bibinfo{pages}{D773} (\bibinfo{year}{2012}).

\bibitem[{\citenamefont{Amor et~al.}(2016)\citenamefont{Amor, Schaub, Yaliraki,
  and Barahona}}]{Amor2016}
\bibinfo{author}{\bibfnamefont{B.~R.~C.} \bibnamefont{Amor}},
  \bibinfo{author}{\bibfnamefont{M.~T.} \bibnamefont{Schaub}},
  \bibinfo{author}{\bibfnamefont{S.~N.} \bibnamefont{Yaliraki}},
  \bibnamefont{and} \bibinfo{author}{\bibfnamefont{M.}~\bibnamefont{Barahona}},
  \bibinfo{journal}{Nature Communications} \textbf{\bibinfo{volume}{7}}
  (\bibinfo{year}{2016}).

\bibitem[{\citenamefont{Triki et~al.}(2018)\citenamefont{Triki, Billot,
  Visseaux, Descamps, Flatters, Camproux, and Regad}}]{triki2018exploration}
\bibinfo{author}{\bibfnamefont{D.}~\bibnamefont{Triki}},
  \bibinfo{author}{\bibfnamefont{T.}~\bibnamefont{Billot}},
  \bibinfo{author}{\bibfnamefont{B.}~\bibnamefont{Visseaux}},
  \bibinfo{author}{\bibfnamefont{D.}~\bibnamefont{Descamps}},
  \bibinfo{author}{\bibfnamefont{D.}~\bibnamefont{Flatters}},
  \bibinfo{author}{\bibfnamefont{A.-C.} \bibnamefont{Camproux}},
  \bibnamefont{and} \bibinfo{author}{\bibfnamefont{L.}~\bibnamefont{Regad}},
  \bibinfo{journal}{Scientific reports} \textbf{\bibinfo{volume}{8}},
  \bibinfo{pages}{5789} (\bibinfo{year}{2018}).

\bibitem[{\citenamefont{Phillips et~al.}(2005)\citenamefont{Phillips, Braun,
  Wang, Gumbart, Tajkhorshid, Villa, Chipot, Skeel, Kale, and
  Schulten}}]{pmid16222654}
\bibinfo{author}{\bibfnamefont{J.~C.} \bibnamefont{Phillips}},
  \bibinfo{author}{\bibfnamefont{R.}~\bibnamefont{Braun}},
  \bibinfo{author}{\bibfnamefont{W.}~\bibnamefont{Wang}},
  \bibinfo{author}{\bibfnamefont{J.}~\bibnamefont{Gumbart}},
  \bibinfo{author}{\bibfnamefont{E.}~\bibnamefont{Tajkhorshid}},
  \bibinfo{author}{\bibfnamefont{E.}~\bibnamefont{Villa}},
  \bibinfo{author}{\bibfnamefont{C.}~\bibnamefont{Chipot}},
  \bibinfo{author}{\bibfnamefont{R.~D.} \bibnamefont{Skeel}},
  \bibinfo{author}{\bibfnamefont{L.}~\bibnamefont{Kale}}, \bibnamefont{and}
  \bibinfo{author}{\bibfnamefont{K.}~\bibnamefont{Schulten}},
  \bibinfo{journal}{J Comput Chem} \textbf{\bibinfo{volume}{26}},
  \bibinfo{pages}{1781} (\bibinfo{year}{2005}).

\bibitem[{\citenamefont{Vanommeslaeghe and MacKerell}(2012)}]{pmid23146088}
\bibinfo{author}{\bibfnamefont{K.}~\bibnamefont{Vanommeslaeghe}}
  \bibnamefont{and} \bibinfo{author}{\bibfnamefont{A.~D.}
  \bibnamefont{MacKerell}}, \bibinfo{journal}{J Chem Inf Model}
  \textbf{\bibinfo{volume}{52}}, \bibinfo{pages}{3144} (\bibinfo{year}{2012}).

\bibitem[{\citenamefont{Grewal and Roy}(2015)}]{pmid26216263}
\bibinfo{author}{\bibfnamefont{R.~K.} \bibnamefont{Grewal}} \bibnamefont{and}
  \bibinfo{author}{\bibfnamefont{S.}~\bibnamefont{Roy}},
  \bibinfo{journal}{Protein Pept. Lett.} \textbf{\bibinfo{volume}{22}},
  \bibinfo{pages}{923} (\bibinfo{year}{2015}).

\bibitem[{\citenamefont{Pastor-Satorras
  et~al.}(2015)\citenamefont{Pastor-Satorras, Castellano, Van~Mieghem, and
  Vespignani}}]{RevModPhys.87.925}
\bibinfo{author}{\bibfnamefont{R.}~\bibnamefont{Pastor-Satorras}},
  \bibinfo{author}{\bibfnamefont{C.}~\bibnamefont{Castellano}},
  \bibinfo{author}{\bibfnamefont{P.}~\bibnamefont{Van~Mieghem}},
  \bibnamefont{and}
  \bibinfo{author}{\bibfnamefont{A.}~\bibnamefont{Vespignani}},
  \bibinfo{journal}{Rev. Mod. Phys.} \textbf{\bibinfo{volume}{87}},
  \bibinfo{pages}{925} (\bibinfo{year}{2015}).

\bibitem[{\citenamefont{Allen}(1994)}]{allen1994some}
\bibinfo{author}{\bibfnamefont{L.~J.} \bibnamefont{Allen}},
  \bibinfo{journal}{Mathematical biosciences} \textbf{\bibinfo{volume}{124}},
  \bibinfo{pages}{83} (\bibinfo{year}{1994}).

\bibitem[{\citenamefont{{R Core Team}}(2013)}]{rr}
\bibinfo{author}{\bibnamefont{{R Core Team}}}, \emph{\bibinfo{title}{R: A
  Language and Environment for Statistical Computing}},
  \bibinfo{organization}{R Foundation for Statistical Computing},
  \bibinfo{address}{Vienna, Austria} (\bibinfo{year}{2013}),
  \bibinfo{note}{{ISBN} 3-900051-07-0},
  \urlprefix\url{http://www.R-project.org/}.

\bibitem[{\citenamefont{Amadei et~al.}(2017)\citenamefont{Amadei, Galdo, and
  D'Abramo}}]{Amadei_2017}
\bibinfo{author}{\bibfnamefont{A.}~\bibnamefont{Amadei}},
  \bibinfo{author}{\bibfnamefont{S.~D.} \bibnamefont{Galdo}}, \bibnamefont{and}
  \bibinfo{author}{\bibfnamefont{M.}~\bibnamefont{D'Abramo}},
  \bibinfo{journal}{Journal of Biomolecular Structure and Dynamics} pp.
  \bibinfo{pages}{1--9} (\bibinfo{year}{2017}).

\bibitem[{\citenamefont{Tartaglia et~al.}(2007)\citenamefont{Tartaglia,
  Cavalli, and Vendruscolo}}]{Tartaglia_2007}
\bibinfo{author}{\bibfnamefont{G.~G.} \bibnamefont{Tartaglia}},
  \bibinfo{author}{\bibfnamefont{A.}~\bibnamefont{Cavalli}}, \bibnamefont{and}
  \bibinfo{author}{\bibfnamefont{M.}~\bibnamefont{Vendruscolo}},
  \bibinfo{journal}{Structure} \textbf{\bibinfo{volume}{15}},
  \bibinfo{pages}{139} (\bibinfo{year}{2007}).

\bibitem[{\citenamefont{Vishveshwara et~al.}(2002)\citenamefont{Vishveshwara,
  Brinda, and Kannan}}]{VISHVESHWARA2002}
\bibinfo{author}{\bibfnamefont{S.}~\bibnamefont{Vishveshwara}},
  \bibinfo{author}{\bibfnamefont{K.~V.} \bibnamefont{Brinda}},
  \bibnamefont{and} \bibinfo{author}{\bibfnamefont{N.}~\bibnamefont{Kannan}},
  \bibinfo{journal}{Journal of Theoretical and Computational Chemistry}
  \textbf{\bibinfo{volume}{01}}, \bibinfo{pages}{187} (\bibinfo{year}{2002}).

\bibitem[{\citenamefont{Vijayabaskar and
  Vishveshwara}(2010)}]{Vijayabaskar2010}
\bibinfo{author}{\bibfnamefont{M.}~\bibnamefont{Vijayabaskar}}
  \bibnamefont{and}
  \bibinfo{author}{\bibfnamefont{S.}~\bibnamefont{Vishveshwara}},
  \bibinfo{journal}{Biophysical Journal} \textbf{\bibinfo{volume}{99}},
  \bibinfo{pages}{3704} (\bibinfo{year}{2010}).

\bibitem[{\citenamefont{Lee et~al.}(2014)\citenamefont{Lee, Wang, Hwang, and
  Tseng}}]{pmid25393107}
\bibinfo{author}{\bibfnamefont{C.~W.} \bibnamefont{Lee}},
  \bibinfo{author}{\bibfnamefont{H.~J.} \bibnamefont{Wang}},
  \bibinfo{author}{\bibfnamefont{J.~K.} \bibnamefont{Hwang}}, \bibnamefont{and}
  \bibinfo{author}{\bibfnamefont{C.~P.} \bibnamefont{Tseng}},
  \bibinfo{journal}{PLoS ONE} \textbf{\bibinfo{volume}{9}},
  \bibinfo{pages}{e112751} (\bibinfo{year}{2014}).

\bibitem[{\citenamefont{Folch et~al.}(2010)\citenamefont{Folch, Dehouck, and
  Rooman}}]{pmid20159163}
\bibinfo{author}{\bibfnamefont{B.}~\bibnamefont{Folch}},
  \bibinfo{author}{\bibfnamefont{Y.}~\bibnamefont{Dehouck}}, \bibnamefont{and}
  \bibinfo{author}{\bibfnamefont{M.}~\bibnamefont{Rooman}},
  \bibinfo{journal}{Biophys. J.} \textbf{\bibinfo{volume}{98}},
  \bibinfo{pages}{667} (\bibinfo{year}{2010}).

\bibitem[{\citenamefont{Folch et~al.}(2008)\citenamefont{Folch, Rooman, and
  Dehouck}}]{pmid18161956}
\bibinfo{author}{\bibfnamefont{B.}~\bibnamefont{Folch}},
  \bibinfo{author}{\bibfnamefont{M.}~\bibnamefont{Rooman}}, \bibnamefont{and}
  \bibinfo{author}{\bibfnamefont{Y.}~\bibnamefont{Dehouck}},
  \bibinfo{journal}{J Chem Inf Model} \textbf{\bibinfo{volume}{48}},
  \bibinfo{pages}{119} (\bibinfo{year}{2008}).

\bibitem[{\citenamefont{Frishman and Argos}(1995)}]{Frishman1995}
\bibinfo{author}{\bibfnamefont{D.}~\bibnamefont{Frishman}} \bibnamefont{and}
  \bibinfo{author}{\bibfnamefont{P.}~\bibnamefont{Argos}},
  \bibinfo{journal}{Proteins: Structure, Function, and Genetics}
  \textbf{\bibinfo{volume}{23}}, \bibinfo{pages}{566} (\bibinfo{year}{1995}).

\bibitem[{\citenamefont{Sengupta and Kundu}(2012)}]{Sengupta2012}
\bibinfo{author}{\bibfnamefont{D.}~\bibnamefont{Sengupta}} \bibnamefont{and}
  \bibinfo{author}{\bibfnamefont{S.}~\bibnamefont{Kundu}},
  \bibinfo{journal}{{BMC} Bioinformatics} \textbf{\bibinfo{volume}{13}}
  (\bibinfo{year}{2012}).

\bibitem[{\citenamefont{B\"{o}de et~al.}(2007)\citenamefont{B\"{o}de,
  Kov{\'{a}}cs, Szalay, Palotai, Korcsm{\'{a}}ros, and Csermely}}]{Bde2007}
\bibinfo{author}{\bibfnamefont{C.}~\bibnamefont{B\"{o}de}},
  \bibinfo{author}{\bibfnamefont{I.~A.} \bibnamefont{Kov{\'{a}}cs}},
  \bibinfo{author}{\bibfnamefont{M.~S.} \bibnamefont{Szalay}},
  \bibinfo{author}{\bibfnamefont{R.}~\bibnamefont{Palotai}},
  \bibinfo{author}{\bibfnamefont{T.}~\bibnamefont{Korcsm{\'{a}}ros}},
  \bibnamefont{and} \bibinfo{author}{\bibfnamefont{P.}~\bibnamefont{Csermely}},
  \bibinfo{journal}{{FEBS} Letters} \textbf{\bibinfo{volume}{581}},
  \bibinfo{pages}{2776} (\bibinfo{year}{2007}).

\end{thebibliography}


\begin{thebibliography}{3}
\expandafter\ifx\csname natexlab\endcsname\relax\def\natexlab#1{#1}\fi
\expandafter\ifx\csname bibnamefont\endcsname\relax
  \def\bibnamefont#1{#1}\fi
\expandafter\ifx\csname bibfnamefont\endcsname\relax
  \def\bibfnamefont#1{#1}\fi
\expandafter\ifx\csname citenamefont\endcsname\relax
  \def\citenamefont#1{#1}\fi
\expandafter\ifx\csname url\endcsname\relax
  \def\url#1{\texttt{#1}}\fi
\expandafter\ifx\csname urlprefix\endcsname\relax\def\urlprefix{URL }\fi
\providecommand{\bibinfo}[2]{#2}
\providecommand{\eprint}[2][]{\url{#2}}

\bibitem[{\citenamefont{Vespignani}(2011)}]{Vespignani2011}
\bibinfo{author}{\bibfnamefont{A.}~\bibnamefont{Vespignani}},
  \bibinfo{journal}{Nature Physics} \textbf{\bibinfo{volume}{8}},
  \bibinfo{pages}{32} (\bibinfo{year}{2011}).

\bibitem[{\citenamefont{Yang and Wang}(2016)}]{Yang2016}
\bibinfo{author}{\bibfnamefont{H.-X.} \bibnamefont{Yang}} \bibnamefont{and}
  \bibinfo{author}{\bibfnamefont{B.-H.} \bibnamefont{Wang}},
  \bibinfo{journal}{Physica A: Statistical Mechanics and its Applications}
  \textbf{\bibinfo{volume}{443}}, \bibinfo{pages}{86} (\bibinfo{year}{2016}).

\bibitem[{\citenamefont{Pastor-Satorras
  et~al.}(2015)\citenamefont{Pastor-Satorras, Castellano, Van~Mieghem, and
  Vespignani}}]{RevModPhys.87.925}
\bibinfo{author}{\bibfnamefont{R.}~\bibnamefont{Pastor-Satorras}},
  \bibinfo{author}{\bibfnamefont{C.}~\bibnamefont{Castellano}},
  \bibinfo{author}{\bibfnamefont{P.}~\bibnamefont{Van~Mieghem}},
  \bibnamefont{and}
  \bibinfo{author}{\bibfnamefont{A.}~\bibnamefont{Vespignani}},
  \bibinfo{journal}{Rev. Mod. Phys.} \textbf{\bibinfo{volume}{87}},
  \bibinfo{pages}{925} (\bibinfo{year}{2015}).

\end{thebibliography}

\end{document}


\font\myfont=cmr12 at 40pt
\title{\huge Supplementary Material  \\
 \large Simulated Epidemics in 3D Protein Structures to Detect Functional Properties\\}

\author{Mattia Miotto\footnote{The authors contributed equally to the present work.}}
\affiliation{Department of Physics, Sapienza University of Rome, Rome, Italy}
\affiliation{Center for Life Nanoscience, Istituto Italiano di Tecnologia,  Rome, Italy}

\author{Lorenzo Di Rienzo\footnotemark[1]}
\affiliation{Department of Physics, Sapienza University of Rome, Rome, Italy}

\author{Pietro Corsi}
\affiliation{University ``Roma Tre'', Department of Science, Rome, Italy}

\author{Giancarlo Ruocco}
\affiliation{Department of Physics, Sapienza University of Rome, Rome, Italy}
\affiliation{Center for Life Nanoscience, Istituto Italiano di Tecnologia,  Rome, Italy}

\author{Domenico Raimondo}
\affiliation{Department of Molecular Medicine, Sapienza University of Rome, Rome, Italy}

\author{Edoardo Milanetti}
\affiliation{Department of Physics, Sapienza University of Rome, Rome, Italy}
\affiliation{Center for Life Nanoscience, Istituto Italiano di Tecnologia,  Rome, Italy}

\keywords{diffusion, information, transmission, thermal stability, proteins, melting temperature, energy, structure}

\newcommand{\beq}{\begin{equation}}
\newcommand{\eeq}{\end{equation}}

\flushbottom
\maketitle

\thispagestyle{empty}
\newpage

\section{Epidemic models}

In the last decades epidemic spreading has seen applications in several fields \cite{Vespignani2011, Yang2016} thanks to the growth of network sciences. Starting from the spread of real diseases and going to news diffusion in social networks, the epidemic models gives a measure about the information diffusion within either the whole network or from a particular node to any other. Epidemic modeling describes the dynamical evolution of the contagion process within a population. A node (or individual) is said susceptible (S) when it is "healthy", infected (I) when the contagion is transmitted by an adjacent node and recovered (R) when it is immunized and hence is safe from other infections. In order to understand the time evolution of the density of infected individuals we have to define the basic processes that rule the transition of individuals from one of the states early described to another. 
A description of \textbf{SIS} (Susceptible-Infected- Susceptible), \textbf{SIR} (Susceptible-Infected-Recovered) and \textbf{SI} (Susceptible-Infected) models with the classical epidemiology \cite{RevModPhys.87.925} is given. Approaches based on mean-field theory are also available, but we are more interested in a clear explanation of the algorithms and their applications rather than their theoretical treatment. An epidemic process can be seen as a stochastic reaction-diffusion process (van Kampen, 1981) in which individuals belonging to the different states (i.e. S, I or R) evolve according to a given set of mutual interaction rules. These represent the possible transitions between the states that can be specified by stoichiometric equations, indeed we can adopt the reaction-diffusion formalism to describe simple epidemic models. The SIS model thus obeys the reactions
\begin{equation} \label{eq:sis1}
S + I \overset {\beta}{\to} 2I
\end{equation}
\begin{equation} \label{eq:sis2}
I \overset {\mu}{\to} S
\end{equation}
where $\beta$ and $\mu$ are transition rates for infection and recovery respectively, that is the probabilities of a node to be infected by an adjacent node or to be recovered ad hence removed (or immunized) by the network. In this model infection can be sustained forever for sufficiently large $\beta$ or small $\mu$. The SIR is instead a three-states model coupled by the reactions
\begin{equation} \label{eq:sir1}
S + I\overset {\beta}{\to} 2I
\end{equation}
\begin{equation}\label{eq:sir2}
I \overset {\mu}{\to} R.
\end{equation}
For any value of $\beta$ and $\mu$, the SIR process will always extinguish after infecting a given population density. A useful variant is the SI model, which only considers the first transition in SIS (Eq. \ref{eq:sis1}) and SIR (Eq. \ref{eq:sir1}), that is nodes become infected and never leave this state. In this way the process ends when the entire population will be infected. While the SI model is a strong simplification (valid only in cases where the time scale of recovery is much larger than the time scale of infection), it approximates the initial time evolution of both SIS and SIR dynamics. The classical description of epidemic dynamics is based on taking the continuous-time limit of difference equations for the evolution of the average number of nodes in each state.

This deterministic approach relies on the homogeneous mixing approximation, which assumes that the nodes in the population are well mixed and interact with each other completely at random, in such a way that each member in a compartment is treated similarly and indistinguishably from the others in that same compartment.

Under this approximation, information about the state of the epidemics is fully expressed with the node density $\rho_\alpha = N_\alpha/N$, where $N$ is the population size and $\alpha$ is the state.

 The time evolution of the epidemics is described by deterministic differential equations, which are constructed applying the law of mass action, stating that the average change in the population density of each compartment due to interactions is given by the product of the force of infection times the average population density (Hethcote, 2000).
The deterministic equations for the SIR and SIS processes are obtained by applying the law of mass action and read as
\begin{equation}
SIS
  \begin{cases}
    \frac{d\rho^I}{dt} = \beta \rho^I \rho^S - \mu \rho^I \\
    \\
    \frac{d\rho^S}{dt} = - \beta \rho^I \rho^S + \mu \rho^I \\
    \\
    \rho^S = 1 - \rho^I \hspace{0.5 cm} $(normalization condition)$
  \end{cases}
\end{equation}
\begin{equation}
SIR
  \begin{cases}
   \frac{d\rho^I}{dt} = \beta \rho^I \rho^S - \mu \rho^I \\
   \\
   \frac{d\rho^S}{dt} = - \beta \rho^I \rho^S \\
   \\
   \rho^R = 1 - \rho^S - \rho^I  \hspace{0.2 cm} $(normalization condition)$
  \end{cases}
\end{equation}
If we consider the limit $\rho^I \simeq 0$, generally valid at the early stage of the epidemic, we can linearize the above equations obtaining for both the SIS and SIR models the simple equation
\begin{equation}
\frac{d\rho^I}{dt} \simeq (\beta - \mu) \rho^I
\end{equation}
whose solution
\begin{equation}
\rho^I(t) \simeq \rho^I(0) e^{(\beta - \mu)t}
\end{equation}
represents the early time evolution. last Equation illustrates one of the key concepts in the classical theoretical analysis of epidemic models. The number of infectious individuals grows exponentially if 
\begin{equation}
\beta - \mu > 0    \Rightarrow    R_0 = \frac{\beta}{\mu} > 1
\end{equation}
where we have defined the basic reproduction number $R_0$ as the average number of secondary infections caused by a primary case introduced in a fully susceptible population (Anderson and May, 1992). This result allows to define the concept of epidemic threshold: only if $R_0 > 1$ (i.e. if a single infected individual generates on average more than one secondary infection) an infective agent can cause an outbreak of a finite relative size (in SIR-like models) or lead to a steady state with a finite average density of infected individuals, corresponding to an endemic state (in SIS-like models). If $R_0 < 1$ (i.e. if a single infected individual generates less than one secondary infection), the relative size of the epidemics is negligibly small, vanishing in the thermodynamic limit of an infinite population (in SIR-like models) or leading to a unique steady state with all individuals healthy (in SIS-like models).

\section{Datasets}

\begin{table*}[]
    \centering
    \begin{tabular}{||c|c|c|c||}
\hline
\hline
Couple     &     Thermostable&     Mesostable &  Active Site\\ 
\hline 
\hline
1      &     1FJQ     &     1NPC& Tyr157; Asp226; His231; His142 A; His146 A: Glu 166 A; Glu 143\\ 
\hline 
2      &     3PFK     &     2PFK& Asp127; Gly 11; Arg72; Thr125; Arg171\\ 
\hline 
3      &     1RIL     &     2RN2& Asp10; Asp70; Glu48; His124; Asp134\\ 
\hline 
4      &     1BMD     &     4MDH& His186; Asp158\\ 
\hline 
5      &     2PRD     &     1JFD& ---\\ 
\hline 
6      &     1PHP     &     3PGK& ---\\ 
\hline 
7       &     1THM     &     1ST3& ---\\ 
\hline 
8      &     1EBD     &     1LVL& ---\\ 
\hline 
9     &     1BTM     &     1TIM& Gly171; Ser211; Lys13; Glu97; Asn11; His95; Glu165\\ 
\hline 
10     &     1IQZ     &     1DUR& ---\\ 
\hline 
11     &     1VJW     &     1FCA& ---\\ 
\hline 
12     &     1XYZ     &     2EXO& Glu127; Asp235; Asn169; Glu233; His205\\ 
\hline 
13     &     1CAA     &     1IRO& ---\\ 
\hline 
14     &     2GD1     &     4GPD& ---\\
\hline 
15     &     1TIB     &     1LGY& ---\\ 
\hline 
16     &     1ZIP     &     2AKY& ---\\ 
\hline 
17     &     1AIS     &     1VOL& ---\\ 
\hline 
18     &     1FFH     &     1FTS& ---\\ 
\hline 
19     &     1OBR     &     2CTC& His69; Glu72; His196; Arg127; Glu270\\ 
\hline 
20     &     1PHN     &     1CPC& ---\\ 
\hline 
21     &     1BLI     &     1HVX& ---\\ 
\hline 
32     &     1TMY     &     3CHY& ---\\ 
\hline 
23     &     1AYG     &     2PAC& ---\\ 
\hline 
24     &     1GHS     &     1GHR& Glu231; Glu279; Lys282; Glu288\\ 
\hline 
25     &     1BVU     &     1HRD& Asp165; Lys125\\ 
\hline 
26     &     1CIU     &     1CGT& ---\\ 
\hline 
27     &     1OSI     &     1CM7& ---\\ 
\hline 
28     &     3MDS     &     1QNM& ---\\ 
\hline 
29     &     1XGS     &     1MAT& His161; Asp82; Asp93; His153; Glu187;Glu280; Glu187\\ 
\hline 
30     &     1DL3     &     1PII& Cys7; Asp126\\ 
\hline 
31     &     1IGS     &     1PII& Asn180, Ser211; Glu51; Lys110; Glu159; Glu210; Lys53\\ 
\hline 
32     &     1BXB     &     1QT1& ---\\
\hline
\hline
    \end{tabular}
    \caption{Table of the 32 couples of proteins of the Thermal dataset, collected from the literature. The proteins corresponding to the rows of the table with the annotated active site are enzymes and constitute the Enzyme dataset.}
    \label{tab:tm}
\end{table*}

\begin{figure*}[!]
\centering
\includegraphics[width=\textwidth]{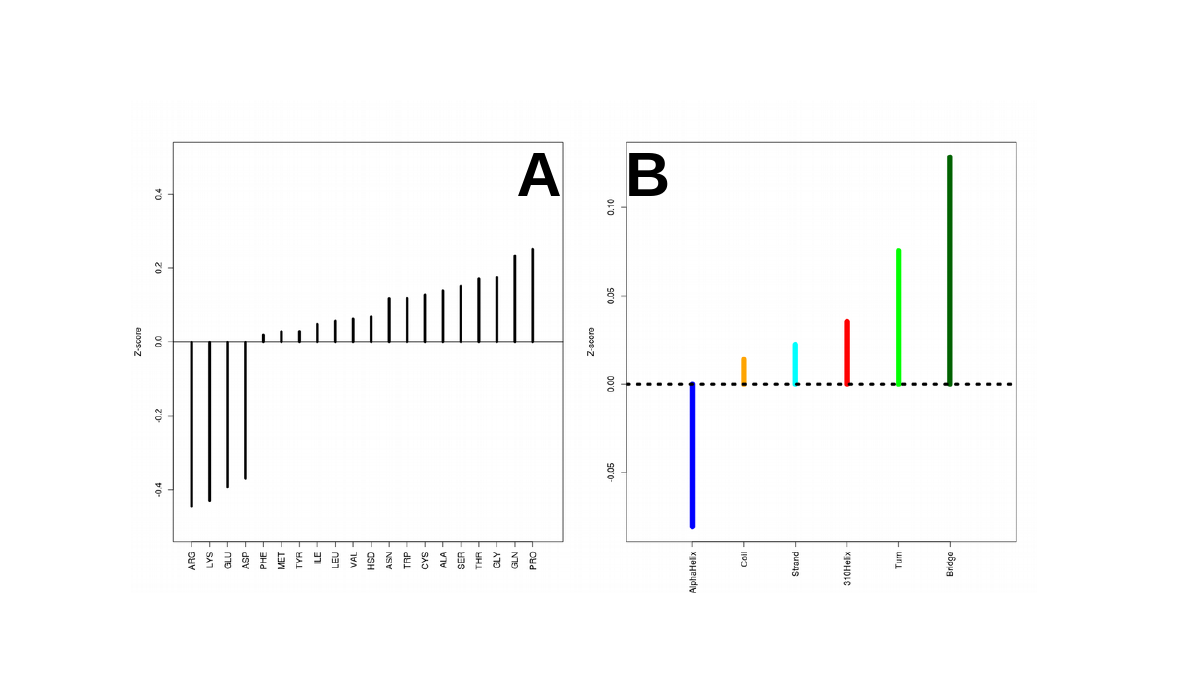}
\caption{\textbf{A)} The mean $t_r$ for each of the 20 amino acids in all proteins of thermostable dataset. \textbf{B)} The mean $t_r$ for each type of secondary structure, obtained using all proteins of thermostable dataset.}
\label{fig:S1}
\end{figure*}

\begin{figure*}[!]
\centering
\includegraphics[width=\textwidth]{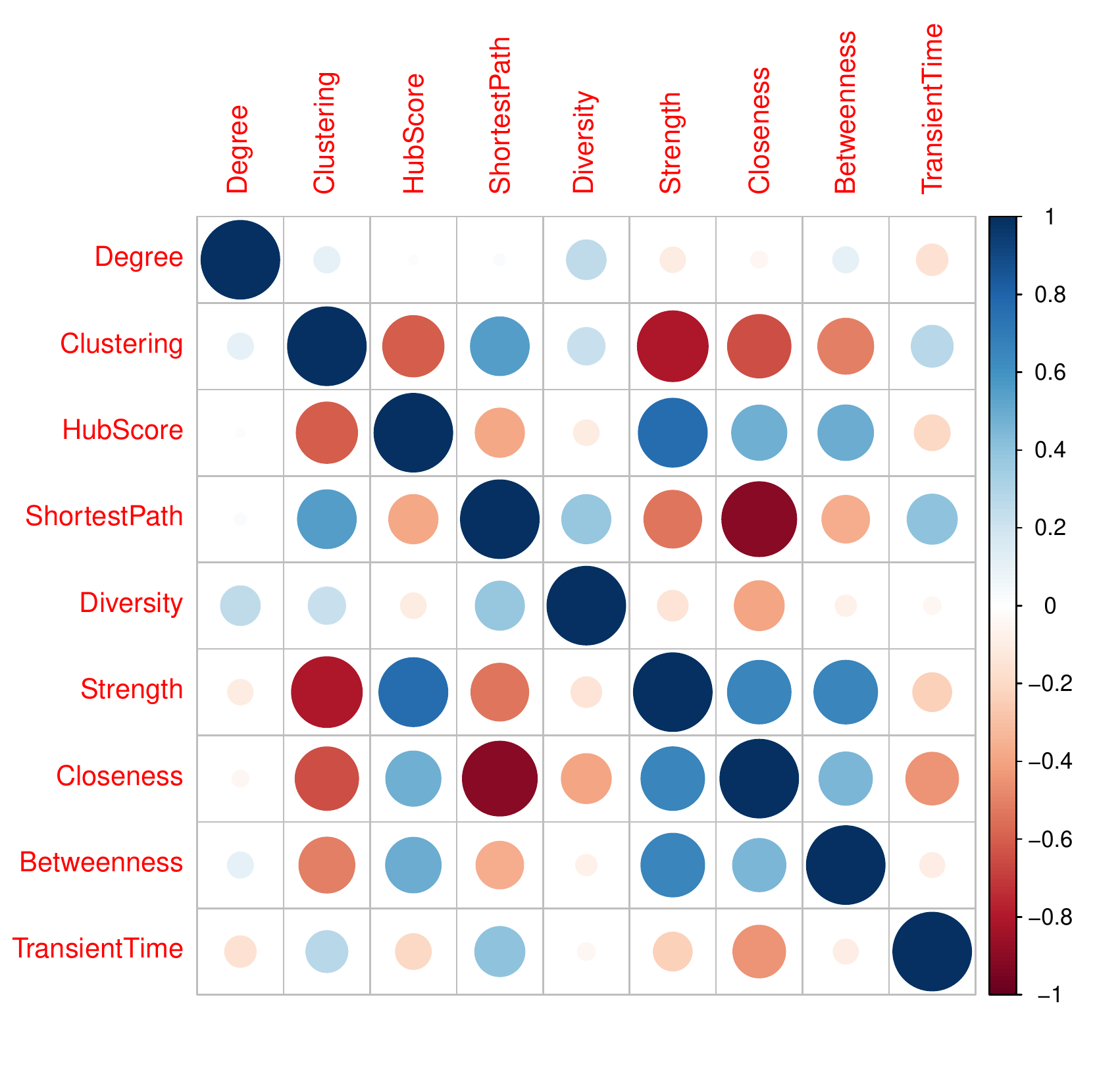}
\caption{Correlation coefficients between common network node descriptors and diffusion transient time, $t_i^\star$.}
\label{fig:S2}
\end{figure*}


\flushbottom
\clearpage
\bibliography{biblioThermo.bib}


